# A Link Clustering Based Approach for Clustering Categorical Data[*]

Zengyou He, Xiaofei Xu, Shengchun Deng

Department of Computer Science and Engineering, Harbin Institute of Technology,

92 West Dazhi Street, P.O Box 315, Harbin 150001, P. R. China

zengyouhe@yahoo.com, {xiaofei, dsc}@hit.edu.cn

**Abstract** Categorical data clustering (CDC) and link clustering (LC) have been considered as separate research and application areas. The main focus of this paper is to investigate the commonalities between these two problems and the uses of these commonalities for the creation of new clustering algorithms for categorical data based on cross-fertilization between the two disjoint research fields. More precisely, we formally transform the CDC problem into an LC problem, and apply LC approach for clustering categorical data. Experimental results on real datasets show that LC based clustering method is competitive with existing CDC algorithms with respect to clustering accuracy.

**Keywords** Clustering, Categorical Data, Link Clustering, Data Mining

## 1. Introduction

Clustering typically groups data into sets in such a way that the intra-cluster similarity is maximized while the inter-cluster similarity is minimized. The clustering technique has been extensively studied in many fields such as pattern recognition [1], customer segmentation [2], similarity search [3] and trend analysis [4].

Most previous clustering algorithms focus on numerical data whose inherent geometric properties can be exploited naturally to define distance functions between data points. However, much of the data existed in the databases is categorical, where attribute values can't be naturally ordered as numerical values. An example of categorical attribute is *shape* whose values include *circle*, *rectangle*, *ellipse*, etc. Due to the special properties of categorical attributes, the clustering of categorical data seems more complicated than that of numerical data. A few algorithms have been proposed in recent years for clustering categorical data [5-24].

Co-occurrence data is an increasingly important and abundant source of data. This type of data has long been important in the social sciences, marketing, and the government intelligence community [25]. In one general form the input consists of a series of *links*. Each link is a set of entities that have been joined by some event or relation. Link Clustering (LC) attempts to find underlying groups of entities given link and demographic data [25,26]. Kubica et al. [25,26] conducted some research efforts on LC.

Until recently, CDC and LC have been considered as separate research and application areas.

---

[*] This work was supported by the High Technology Research and Development Program of China (No. 2002AA413310, No. 2003AA4Z2170, No. 2003AA413021) and the IBM SUR Research Fund.

The starting point in this paper is the observation of some key underlying similarities between these two different areas. This observation makes possible the study of CDC problem from a LC perspective. This different perspective may enable a better understanding of the CDC algorithms and help in devising improved or hybrid versions by combining elements from areas that would otherwise be considered incompatible. That is, our first contribution is the exploration of underlying properties, similarities and differences between CDC and LC, which creates the basis for the proposal of LC based clustering algorithms for categorical data.

Our second contribution is the direct adaptation and use of LC methodology for clustering categorical data. We formally define the CDC problem as an LC problem, and apply LC approach for clustering categorical data. Our experimental results show the new categorical data clustering methods to achieve better clustering accuracy than previous algorithms, which confirms our intuition that LC approaches and CDC methods can be used interchangeably. Furthermore, the idea of linking LC and CDC will enable a problem at hand to be solved through either way. Thus, improvements can be achieved in both domains.

The remainder of this paper is organized as follows. Section 2 presents a critical review on related work. Section 3 creates an interesting view on the underlying properties, similarities and differences between CDC and LC. In Section 4, we define the CDC problem as an optimization problem and describe the LC based algorithms for clustering categorical data. Experimental results are given in Section 5 and Section 6 concludes the paper.

## 2. Related Work

### 2.1 Clustering Categorical Data

A few algorithms have been proposed in recent years for clustering categorical data [5-24]. In [5], the problem of clustering customer transactions in a market database is addressed. STIRR, an iterative algorithm based on non-linear dynamical systems is presented in [6]. The approach used in [6] can be mapped to a certain type of non-linear systems. If the dynamical system converges, the categorical databases can be clustered. Another recent research [7] shows that the known dynamical systems cannot guarantee convergence, and proposes a revised dynamical system in which convergence can be guaranteed.

K-modes, an algorithm extending the *k*-means paradigm to categorical domain is introduced in [8,9]. New dissimilarity measures to deal with categorical data is conducted to replace means with modes, and a frequency based method is used to update modes in the clustering process to minimize the clustering cost function. Based on *k*-modes algorithm, [10] proposes an adapted mixture model for categorical data, which gives a probabilistic interpretation of the criterion optimized by the *k*-modes algorithm. A fuzzy *k*-modes algorithm is presented in [11] and tabu search technique is applied in [12] to improve fuzzy *k*-modes algorithm. An iterative initial-points refinement algorithm for categorical data is presented in [13]. The work in [23] can be considered as the extensions of *k*-modes algorithm to transaction domain.

In [14], the authors introduce a novel formalization of a cluster for categorical data by generalizing a definition of cluster for numerical data. A fast summarization based algorithm, CACTUS, is presented. CACTUS consists of three phases: *summarization*, *clustering*, and *validation*.

ROCK, an adaptation of an agglomerative hierarchical clustering algorithm, is introduced in [15]. This algorithm starts by assigning each tuple to a separated cluster, and then clusters are merged repeatedly according to the closeness between clusters. The closeness between clusters is defined as the sum of the number of "links" between all pairs of tuples, where the number of "links" is computed as the number of common neighbors between two tuples.

In [16], the authors propose the notion of *large item*. An item is *large* in a cluster of transactions if it is contained in a user specified fraction of transactions in that cluster. An allocation and refinement strategy, which has been adopted in partitioning algorithms such as *k*-means, is used to cluster transactions by minimizing the criteria function defined with the notion of large item. Following the large item method in [16], a new measurement, called the small-large ratio is proposed and utilized to perform the clustering [17]. In [18], the authors consider the item taxonomy in performing cluster analysis. While the work [19] proposes an algorithm based on "caucus", which is fine-partitioned demographic groups that is based the purchase features of customers.

Squeezer, a one-pass algorithm is proposed in [20]. *Squeezer* repeatedly read tuples from dataset one by one. When the first tuple arrives, it forms a cluster alone. The consequent tuples are either put into an existing cluster or rejected by all existing clusters to form a new cluster by given similarity function.

COOLCAT, an entropy-based algorithm for categorical clustering, is proposed in [21]. Starting from a heuristic method of increasing the height-to-width ratio of the cluster histogram, the authors in [22] develop the CLOPE algorithm. [24] introduce a distance measure between partitions based on the notion of generalized conditional entropy and a genetic algorithm approach is utilized for discovering the median partition.

## 2.2 Link Clustering

Kubica *et. al.* presented the group detection algorithm (GDA) – an algorithm for finding underlying groupings of entities from co-occurrence data [26]. This algorithm is based on a probabilistic generative model and produces coherent groups that are consistent with prior knowledge.

In [26], Kubica *et. al.* present k-groups - an algorithm that uses an approach similar to that of k-means (hard clustering and localized updates) to significantly accelerate the discovery of the underlying groups while retaining GDA's probabilistic model.

## 3. A Unified View on CDC and LC

The researches on CDC and LC have been conducted in parallel. Our goal in this section is to argue that a unified view can be built for the CDC problem and CE problem, hence, CDC problem can be solved with existing LC algorithms.

Clustering aims at discovering groups and identifying interesting patterns in a dataset. We call a particular clustering algorithm with a specific view of the data a *clusterer*. Each clusterer outputs a *clustering* or *labeling*, comprising the group labels for some or all objects. Link Clustering (LC) attempts to find underlying groups of entities given link and demographic data

[25,26].

Consider the dataset $X = \{x_1, x_2 \ldots x_n\}$ be a set of objects described by $r$ categorical attributes, $A_1, \ldots, A_r$ with domains $D_1, \ldots, D_r$ respectively. Two rows $t$ and $u$ are equivalent with respect to a given attribute $A_i$ if $t[A_i] = u[A_i]$. Informally, we can say that $t$ and $u$ are "*linked*" by their common attribute values on $A_i$.

Furthermore, any attribute $A_i$ partitions the rows of the relation into equivalence classes. We denote the *equivalence class* of a row $t \in X$ with respect to a given attribute $A_i$ by $[t]_{A_i}$, i.e., $[t]_{A_i} = \{u \in X \mid t[A_i] = u[A_i]\}$, the set $\pi_{A_i} = \{[t]_{A_i} \mid t \in X\}$ of equivalence classes is a partition of X under $A_i$. That is, $\pi_{A_i}$ is a collection of disjoint sets (equivalence classes) of rows, such that each set has a unique value for the attribute $A_i$, and the union of the sets equals the $X$. From the viewpoint of LC, each set in $\pi_{A_i}$ is a *link*. Therefore, we can employ existing LC algorithms for solving the CDC problem.

More formally, for the LC problem, the input data is assumed to contain $N_P$ unique entities. We denote the set of all entities as $\xi$ and a single entity as $e_i \in \xi$. The input data consists of $N_L$ links, each of which is a subset of entities. A single link is denoted $L_i \subseteq \xi$ and contains $|L_i|$ entities. The entire set of links is denoted as $L_D$. Finally we are attempting to find $K$ groups. We denote a group as $g_i \in \xi$ and the set of all current groups as $G$. Hence, we can transform the CDC problem into a LC problem as follows:

Let $\xi = \{1, 2, \ldots, n\}$,

$N_P = |X| = n$,

$L_D = \{\lambda \in \pi_{A_i} \mid i = 1, 2, \ldots, r\}$,

$N_L = |L_D|$.

With these transformations, if we can find $K$ groups on $\xi$, we solve the corresponding CDC problem. Similarly, a reverse process for transforming the LC problem into CDC problem could be established. That is, CDC problem and LC problem are equivalent, and algorithms developed in both domains can be used interchangeably.

**Example 1** Table 1 shows a categorical table with 10 records, each described by 2 categorical attributes. Only considering "*Attribute 1*", we can get the partition {(1,2,5,7,10), (3,4,6,8,9)}. Similarly, "*Attribute 2*" gives a partition as {(1,4,9), (2,3,10), (5,6,7,8)}. Thus, clustering this categorical data can be transformed as a LC problem as:

$\xi = \{1, 2, \ldots, 10\}$, $N_P = |X| = 10$,

$$L_D = \{\{1,2,5,7,10\},\{3,4,6,8,9\},\{1,4,9\},\{2,3,10\},\{5,6,7,8\}\}, \ N_L = 5.$$

Table 1 Sample Categorical Data Set

| Record Number | Attribute 1 | Attribute 2 |
|---|---|---|
| 1 | M | A |
| 2 | M | B |
| 3 | F | B |
| 4 | F | A |
| 5 | M | C |
| 6 | F | C |
| 7 | M | C |
| 8 | F | C |
| 9 | F | A |
| 10 | M | B |

## 4. Link Clustering Based Approach

As shown in Section 3, the CDC problem can be regarded as an equivalent LC problem. In this section, we borrow the idea of link clustering [25, 26] to solve CDC problem and describe those LC based algorithms for clustering categorical data.

Kubica *et. al.* [25,26] presented GDA and k-groups algorithms, both of which is based on a probabilistic generative model shown in Figure 1. The model is a Bayesian network where the ovals indicate parameters to be learned from data. The five primary components of this model are:

1. The *demographics data set* (*DD*), which contains all the entities under consideration and their demographic information.

2. The *link data set* (*LD*), which is just a set of records specifying observed links.

3. The *chart* (*CH*), which indicates which entities belong to which groups. This data structure can be pictured as a sparse $N_P \times K$ matrix where the columns indicate the groups and the rows indicate the entities. Unlike traditional clustering models, an entity can simultaneously be a full member of several groups (have a "one" in several columns);

4. The *demographic model* (*DM*), which defines a recipe for placing a entity in a group based on demographic information; and

5. The *link model* (*LM*), which defines a recipe for link generation. The link model contains two parameters for each link type: the probability a link of that type is completely random (innocent), $P_I$, and the probability that an entity in a link of that type is noise (random), $P_R$. These probabilities can be learned or specified by the user.

The above components provide a recipe for generating the data from the model. Using the decomposition provided by the Bayesian network in Figure 1, the actual maximum likelihood estimation is accomplished by using noisy hill climbing, a heuristic optimization method. The goal is to find the chart, demographics model, and possibly link model that maximize the likelihood as given by the entire Bayesian network in Figure 1. This is done by trying different moves (adding an entity to a group or removing an entity from a group) and evaluating the change in the loglikelihood.

For more details, the reader should refer to [25,26].

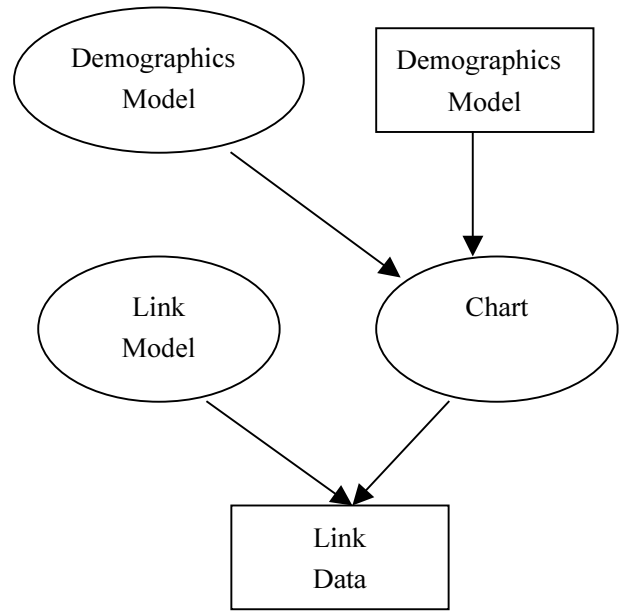

**Fig. 1**: Probabilistic model of group membership and link generation [25].

## 5. Experimental Results

A comprehensive performance study has been conducted to evaluate our method. In this section, we describe those experiments and the results. We ran our algorithm on real-life datasets obtained from the UCI Machine Learning Repository [27] to test its clustering performance against other algorithms.

### 5.1 Real Life Datasets and Evaluation Method

We experimented with four real-life datasets: the Mushroom dataset, and the Cancer dataset, which were obtained from the UCI Machine Learning Repository [27]. Now we will give a brief introduction about these datasets.

- ✓ **The Mushroom Dataset:** It has 22 attributes and 8124 records. Each record represents physical characteristics of a single mushroom. A classification label of poisonous or edible is provided with each record. The numbers of edible and poisonous mushrooms in the dataset are 4208 and 3916, respectively.
- ✓ **Wisconsin Breast Cancer Data[1]:** It has 699 instances with 9 attributes. Each record is

---

[1] We use a dataset that is slightly different from its original format in UCI Machine Learning Repository, which has 683 instances with 444 benign records and 239 *malignant* records. It is public available at:

labeled as *benign* (458 or 65.5%) or *malignant* (241 or 34.5%). In our literature, all attributes are considered categorical with values 1,2, …, 10.

Validating clustering results is a non-trivial task. In the presence of true labels, as in the case of the data sets we used, we can form a confusion matrix to measure the effectiveness of the algorithm. Each entry (*i*, *j*) in the confusion matrix represents the number of objects in cluster *i* that belong to true class *j*. For an objective evaluation measure we use the clustering accuracy for measuring the clustering results was computed as follows. Given the final number of clusters, *k*, clustering accuracy *r* was defined as: $r = \frac{\sum_{i=1}^{k} a_i}{n}$, where *n* is the number of records in the dataset, $a_i$ is the number of instances occurring in both cluster *i* and its corresponding class, which had the maximal value. In other words, $a_i$ is the number of records with the class label that dominates cluster *i*. Consequently, the clustering error is defined as $e = 1 - r$.

## 5.2 Experiment Design

We studied the clustering found by three algorithms, our algorithm denoted as LCBCDC (**L**ink **C**lustering **B**ased **C**ategorical **D**ata **C**lustering), the *Squeezer* algorithm introduced in [20] and the *k*-modes algorithm proposed in [8,9]. It has been demonstrated that the *Squeezer* and *k*-modes algorithm can produce better clustering output than other algorithms in categorical dataset with respect to clustering accuracy. Thus, the two algorithms are selected for the competition.

In all the experiments, except for the number of clusters, all the parameters required by the LCBCDC algorithm are set to be default[2]. The *Squeezer* algorithm requires *only* a similarity threshold as input parameter, so we set this parameter to a proper value to get the desired number of clusters (For the *Squeezer* algorithm, if the output number of clusters is same, the clustering accuracy is almost identical. Hence, we can use *any* similarity threshold value that can make the algorithm get the desired number of clusters). For the *k*-modes algorithm, we use the first *k* distinct records from the data set to construct initial *k* modes.

Moreover, since the clustering results of *k*-modes algorithm and *Squeezer* algorithm are fixed for a particular dataset when the parameters are fixed, only one run is used in the two algorithms. The LCBCDC algorithm is a probabilistic algorithm, so its outputs will differ in different runs. However, we observed in the experiments that the clustering result is very stable, so the clustering error of this algorithm is reported with its first run. In summary, we use one run to get the clustering output for all the three algorithms.

## 5.3 Clustering Results on Mushroom Data

For this dataset, we let all the algorithms produce two clusters. Table 2 shows the clusters and

---

http://research.cmis.csiro.au/rohanb/outliers/breast-cancer/brcancerall.dat.
[2] Since our implementation for the *LCBCDC* algorithm is based on the algorithms developed by Kubica *et. al.* [25,26]. So, the readers may refer to Kubica's link clustering algorithms for details. The executable program of Kubica's link clustering algorithms is available at http://www.autonlab.org/autonweb/showSoftware/147/.

class distribution produced by the three algorithms. As Table 2 shows, *k*-modes algorithm and *Squeezer* algorithm have similar performance, while our algorithm performs much better than *k*-modes algorithm and *Squeezer* algorithm.

Table 2: Clustering results on mushroom data

| \multicolumn{3}{c}{Squeezer(Clustering Error: 0.464)} |
|---|---|---|
| *Cluster NO* | *No of Edible* | *No of Poisonous* |
| 1 | 3723 | 3873 |
| 2 | 485 | 43 |
| \multicolumn{3}{c}{*k*-modes(Clustering Error: 0.435)} |
| *Cluster NO* | *No of Edible* | *No of Poisonous* |
| 1 | 1470 | 1856 |
| 2 | 2738 | 2060 |
| \multicolumn{3}{c}{LCBCDC(Clustering Error: 0.139)} |
| *Cluster NO* | *No of Edible* | *No of Poisonous* |
| 1 | 48 | 1768 |
| 2 | 2950 | 712 |

## 5.4 Clustering Results on Cancer Data

Table 3 contrasts the clustering results on cancer dataset. The clusters produced by LCBCDC algorithm covers only 392 records out of 683 records because the rest are treated as outliers or noises. Recalling the class distribution on the cancer dataset, it has 683 instances with 444 *benign* records and 239 *malignant* records. As shown in Table 3, LCBCDC reasonably discards most of those *malignant* records as outliers and reveals an almost *pure* cluster that is made up with *benign* records. From this perspective, LCBCDC is more capable to discover meaningful clusters.

Moreover, LCBCDC algorithm performs much better than *k*-modes algorithm and *Squeezer* algorithm on this data set, which almost reach the 100% clustering accuracy.

Table 3: Clustering results on cancer data

| \multicolumn{3}{c}{Squeezer(Clustering Error: 0.133)} |
|---|---|---|
| *Cluster NO* | *No of Benign* | *No of malignant* |
| 1 | 442 | 89 |
| 2 | 2 | 150 |
| \multicolumn{3}{c}{*k*-modes(Clustering Error: 0.082)} |
| *Cluster NO* | *No of Benign* | *No of malignant* |
| 1 | 439 | 53 |
| 2 | 5 | 186 |
| \multicolumn{3}{c}{LCBCDC(Clustering Error: 0.003)} |
| *Cluster NO* | *No of Benign* | *No of malignant* |
| 1 | 386 | 1 |
| 2 | 0 | 5 |

## 5.5 Summary

One may argue that the results cannot precisely reflect that our method has better performance since not all records are assigned to a cluster in our method. However, it should be noted that in our method, one object also could be assigned to multiple clusters. Hence, from those results, we are confident to claim that our method could provide at least the same level of accuracy as other popular methods.

## 6. Conclusions

Our main contribution in this paper is to explicitly state the equivalence between the CDC problem and LC problem for the first time, and point out that algorithms developed in both domains can be used interchangeably. Moreover, to verify our statement, we formally show how to transform the CDC problem into a LC problem, and apply LC approach for clustering categorical data. Empirical evidences show that our idea is promising in practice.

In the future, we are planning to explore existing categorical data clustering algorithms, or transactions clustering algorithms to cluster large link dataset.